\documentclass[aps,prl,twocolumn,notitlepage,superscriptaddress]{revtex4-1}

\usepackage{bm}
\usepackage{graphicx}
\usepackage{latexsym}
\usepackage{revsymb}
\usepackage{amsmath}
\usepackage{mathtools}
\usepackage[colorlinks=true,citecolor=blue,linkcolor=blue]{hyperref}
\usepackage{bbding}
\usepackage[caption=false]{subfig} % caption=false - justifies and resizes the caption text
\usepackage{float}
\usepackage{subfloat}
\usepackage[section]{placeins}
\usepackage{csquotes}
\usepackage{relsize}
\usepackage{enumitem}
\usepackage[Symbol]{upgreek}
\DeclareGraphicsExtensions{.pdf,.png,.jpg}

\LetLtxMacro{\OldSqrt}{\sqrt}
\newcommand{\ClosedSqrt}[1][\hphantom{3}]{\def\DHLindex{#1}\mathpalette\DHLhksqrt}
\makeatletter
    \newcommand*\bold@name{bold}
    \def\DHLhksqrt#1#2{%
        \setbox0=\hbox{$#1\OldSqrt{#2\,}$}\dimen0=\ht0\relax%
        \advance\dimen0-0.2\ht0\relax% size of the added box is still 0.2 times ht0
        \setbox2=\hbox{\vrule height\ht0 depth -\dimen0}%
        {\hbox{$#1\expandafter\OldSqrt\expandafter[\DHLindex]{#2\,}$}
        \lower\ifx\math@version\bold@name0.6pt\else0.4pt\fi\box2}
    }
    \renewcommand*{\sqrt}[2][\ ]{\ClosedSqrt[\leftroot{-2}\uproot{1}#1]{#2}\kern0.1em} 
\makeatother

\begin{document}

\title{Nanostructured Tube Wakefield Accelerator}

\author{Aakash A. Sahai}
\affiliation{College of Engineering, Design and Computing, University of Colorado, Denver, CO 80204}
\email[corresponding author: ~]{aakash.sahai@gmail.com}

\author{Toshiki Tajima}
\affiliation{Department of Physics \& Astronomy, University of California, Irvine, CA 92697}

\author{Vladimir D. Shiltsev}
\affiliation{Accelerator Research Department, Fermilab, Batavia, IL 60510}

%\author{Peter Taborek}
%\affiliation{Department of Physics \& Astronomy and Applied Physics, University of California, Irvine, CA 92697}

%\author{Gerard Mourou}
%\affiliation{DER-IZEST, Ecole Polytechnique, 91128 Palaiseau, Cedex, France}

% ABSTRACT
\begin{abstract}
%Atomic acceleration principle which propounds unprecedented $\rm TeVm^{-1}$ energy gain using solid-state fields is yet to be realized. In this work it is demonstrated to be realizable  here overcome by introducing 
Unprecedented $\rm TeVm^{-1}$ acceleration gradients are modeled to be realizable using a nonlinear surface crunch-in mode in nanostructured tubes. This mode is realizable using advances in nanofabrication and solid energy density attosecond bunch compression. Three dimensional computational and analytical modeling demonstrates GeV energy gain in sub-millimeter long tubes with effective wall densities $n_{\rm t}\sim10^{22-24}\rm cm^{-3}$ and hundreds of nanometer core radius when driven by submicron near solid electron beams, $n_{\rm b}\sim0.05n_{\rm t}$. Besides the many $\rm TVm^{-1}$ average gradients, strong self-focusing and nanomodulation of the beam which increases its peak density and the wakefield strength also opens up controlled high-energy photon production.
\end{abstract}

%                             % Classification Scheme.
%                              %display desired
%\pacs{52.27.Ep, 52.38.Kd, 52.40.Mj, 52.65.Rr}
\maketitle

%%%%%%%%%%%%%%%%%%%%%%%%%%%%%%%%%%%%%%%%%%%%%%%%%%%%%%%%%%%%%%%%%%%%%
% SECTION - Introduction
%%%%%%%%%%%%%%%%%%%%%%%%%%%%%%%%%%%%%%%%%%%%%%%%%%%%%%%%%%%%%%%%%%%%%
% INTRODUCTION
%\section{Introduction}
Acceleration of particles in crystals which in principle \cite{atomic-accelerator} propounds unprecedented $\rm TVm^{-1}$ fields using solid-state excitations remains unrealizable. However, the emergence of attosecond charged particle bunches and x-ray laser \cite{xray-Mourou} of solid energy density has kindled \cite{sotwa-2019, wakefield-Tajima} the realizability of $\rm TeVm^{-1}$ acceleration gradients using solid-state wakefields \cite{Chen-solid-state, Tajima-crystal-xray}. Solid-state wakefields have been theorized to sustain nanometric collective modes \cite{Bohm-Pines-electron-gas} that yield $\rm TVm^{-1}$ accelerating fields. These fields are many orders of magnitude higher than both the time-tested radio-frequency accelerators as well as the gaseous plasma wakefield acceleration techniques \cite{Tajima-Dawson-Laser-wakefield, Chen-Dawson-Beam-wakefield}. Naturally, a realizable nano wakefield accelerator promises to not only open new horizons for particle colliders \cite{Shiltsev-crystal} but possibly also non-collider paradigms \cite{TeV-chip} towards Planck-scale physics. But, there have been major challenges to its realizability: (i) lack of intense attosecond bunches to excite collective modes in bulk solids; (ii) drawbacks of direct irradiation of bulk solids.

Here we introduce a novel solid-state nonlinear surface or ``crunch-in'' mode \cite{sotwa-2019, crunch-in-regime, crunch-in-regime-ipac, tajima-ushioda-1978, surface-oscillations, solid-surface-mode-classification} to overcome the key challenges to a nano wakefield accelerator. It relies on the convergence of attosecond compression techniques underlying the emergent relativistic intensity x-ray laser and solid density particle bunches \cite{phase-space-gym} with the advances in nanofabrication \cite{nanotubes-Ijima, ALD-review}. These advances allow an intense bunch propagating in a tube with vacuum-like core \cite{nanotubes-Ijima} of hundreds of nanometer radius, $r_t$ and effective wall densities, $n_t\rm\sim10^{22-24}\rm cm^{-3}$ to drive the tube electrons to crunch in to its core \cite{crunch-in-regime}. The strong electrostatic component of this surface wave helps sustain $\rm TVm^{-1}$ fields without direct interaction with ions. The detailed physics of the proposed mode is here elucidated using 3D computational and analytical model.
%supported by an analytical model.  

%Physical processes of the proposed nanometric surface crunch-in mode \cite{crunch-in-regime, solid-surface-mode-classification} which is excited as wakefield of an intense attosecond bunch in nanostructured tubes \cite{nanotubes-Ijima} are elucidated using three dimensional computational model supported by an analytical model. 

%Although a nano x-ray laser is yet to be prototyped, to conceptualize the solid-state crunch-in mode in a nanostructured tube the collective electron density driven by an x-ray pulse is illustrated in Fig.\ref{} (details below). Existing nanometric beam are here used to establish the realizability of an nano wakefield accelerator.

Excitation of the surface crunch-in mode as wakefields in nanostructured tubes is significantly more practical compared to bulk modes in unstructured solids \cite{Chen-solid-state, Tajima-crystal-xray} because nanofabrication allows better control over structure, density, thickness etc. \cite{nanotubes-Ijima, ALD-review}. This further mitigates the adverse effects of direct irradiation of bulk solids. Investigations of commercial fiber-like tubes \cite{nanotubes-Ijima} using a scanning electron microscope reveal a vacuum-like core with a few nm wall-to-core transition. Deposition of porous material  \cite{ALD-review} on the inner tube surface allows tunable effective density as well as other characteristics. Nanofabricated tubes thus allow in-vacuum propagation of the most populated part of the bunch which overcomes critical obstacles such as collisions, emittance degradation \cite{Chen-solid-state}, filamentation \cite{beam-crystal-filamentation} etc.

Hundreds of nm long bunches are here found to be short enough for controlled excitation of the surface crunch-in mode. Using the crunch-in mode to approach the coherence (wavebreaking) limit of collective fields, $E_{wb} \simeq 9.6 ( n_t [10^{22} {\rm cm^{-3}}] )^{1/2} ~ \rm TVm^{-1}$ \cite{Tajima-Dawson-Laser-wakefield} additionally requires near solid density beam, $n_{\rm b} \sim 0.05 n_{\rm t}$. While a few micron nC electron bunches which approach these requirements are used in x-ray free electron lasers \cite{XFEL-bright-beam} (XFEL), near solid submicron bunches are also currently being commissioned (FACET-II facility \cite{phase-space-gym}). Such beams can still drive wavebreaking wakefields ($\sim E_{wb}$) using a self-focussing effect similar to that demonstrated at the recent AWAKE experiment \cite{AWAKE-nature}.

%Nonlinear surface crunch-in mode is here demonstrated to be excited using hundreds of nm long charged particle bunches in tubes with wall electron densities $n_{\rm t}\simeq10^{22-24}\rm cm^{-3}$. Electron beams in operational XFELs approach these requirements with few $\rm\mu m$ long nC bunches \cite{SASSE-FEL-beam}. Near solid-density bunches $n_{\rm b} \gtrsim 0.05 n_{\rm t}$ are demonstrated to excite collective fields of the coherence (wavebreaking) limit, $E_{wb} \simeq 9.6 ( n_t [10^{22} {\rm cm^{-3}}] )^{1/2} ~ \rm TVm^{-1}$ \cite{Tajima-Dawson-Laser-wakefield} using beam nanomodulation, similar to that in CERN AWAKE facility \cite{AWAKE-nature}

Electric fields of solid density charged particle beams approach $\rm TVm^{-1}$. The hundreds of eV potential of solid beams over atomic scales ($\sim$ 10\AA) is thus significantly higher than the few eV electron binding energy in nanostructured materials \cite{work-function-CNT}. Moreover, the so unbound electrons acquire relativistic momenta over atomic scales in the presence of $\rm TVm^{-1}$ beam fields. 

In contrast, solid-state modes based on oscillations of conduction band electron gas (plasmon) have oscillatory velocities just higher than the Fermi velocity, $v_{\rm F}$ \cite{Bohm-Pines-electron-gas}. Moreover, whereas the collisionless nature of conduction band electron oscillations in bulk solid \cite{Bohm-Pines-electron-gas} or on its surface \cite{surface-oscillations} is known \cite{Bloch-Tomonaga}, it could be enhanced in nanostructures which manifest micron level mean free paths \cite{NT-mfp-2007}. Due to these distinctive characteristics, attosecond collective electron dynamics in relativistic excitation of nanostructures approximates that of a collisionless quasi-neutral electron gas in background ionic lattice.

%%%%%%%%%%%%%%%%%%%%%%%%%%%%%%%%%%%%%%%%%
% FIGURE - Tube dynamics of the wakefield
%%%%%%%%%%%%%%%%%%%%%%%%%%%%%%%%%%%%%%%%%
\begin{figure}[!htb]
\centering
   \includegraphics[width=\columnwidth]{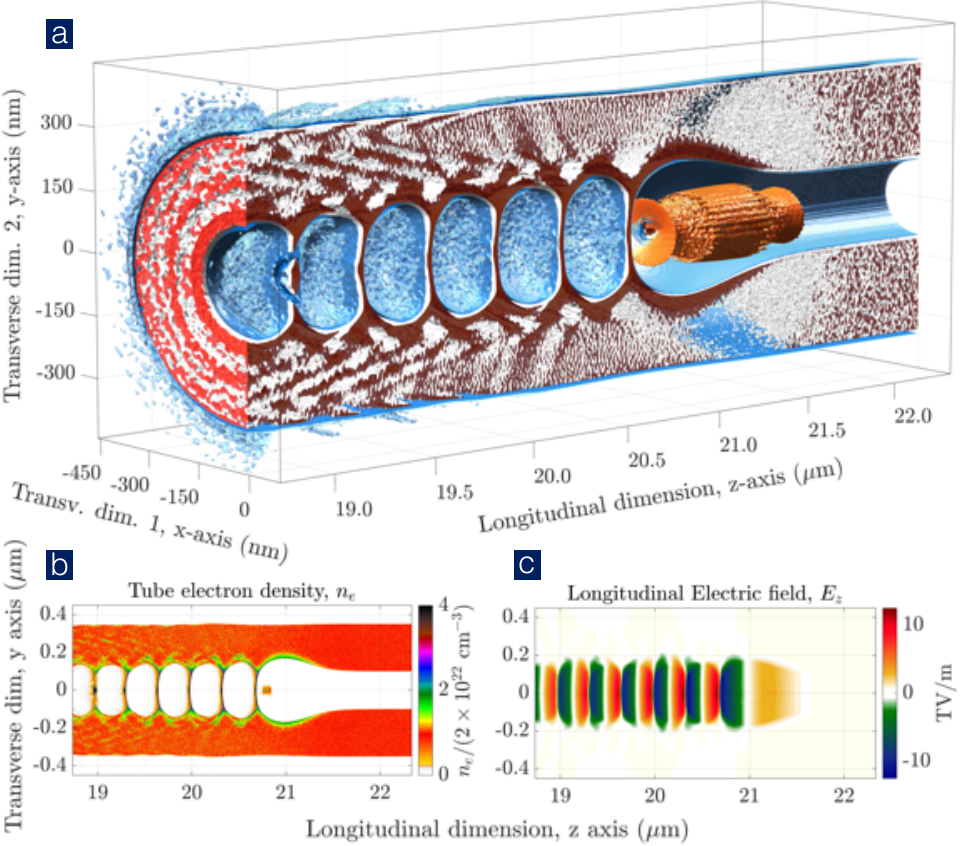}
   \caption{3D PIC simulation (a,b) electron density and (c) longitudinal field profile of crunch-in tube wakefield at around $\rm 20\mu m$ of interaction of a $\sigma_z\rm=400nm$ beam with a nanostructured tube of core radius, $r_t\rm=100nm$.}
\label{fig:3D-wakefields-beam-tube}
\end{figure}

The nonlinear surface mode modeled here thus exhibits several unique features. Firstly, it is a relativistic nonlinear generalization of the surface plasmon polariton (SPP) mode \cite{tajima-ushioda-1978,solid-surface-mode-classification}. Secondly, while the SPP mode is sustained by small-scale surface electron oscillations, here the high oscillation amplitude leads to the crunch in of tube wall electrons deep into its core. Lastly, quintessential solid-state properties such as energy quantization and periodic ion lattice potential become less relevant \cite{xray-wakefield}.

In a quasineutral ideal electron gas of density $n_0$, effective excitation of wakefields necessitates bunch compression of the order of $\omega_{pe}^{-1} = [4\pi n_0e^2m_e^{-1}]^{-1/2} = {\rm 177} (n_0 {\rm [10^{22} cm^{-3}] )^{-1/2}} ~ {\rm attoseconds}$ and $\lambda_{pe} = 2\pi c\omega_{pe}^{-1} = {\rm 333 (}n_0 {\rm [10^{22} cm^{-3}])^{-1/2} ~ nm}$ \cite{Bohm-Pines-electron-gas}. Access to wavebreaking fields, $E_{wb}[n_0]=m_ec\omega_{pe}e^{-1}$ also requires sufficient energy density which further necessitates a minimum number of particles in the ultrashort bunch.

The thrust towards attosecond photon \cite{xray-Mourou} and particle \cite{phase-space-gym} compression favors effective excitation of solid-state wakefields. A precedent has been set by sub picosecond, micron scale (i) chirped pulse amplified \cite{morou-CPA} optical lasers and (ii) compressed \cite{phase-space-gym} or plasma modulated \cite{AWAKE-nature} particle bunches matched to the gaseous plasmas with $n_0 \simeq 10^{16-18}{\rm cm^{-3}}$ ($\omega_{pe}^{-1} \simeq {\rm 10fs}$) that have made possible the demonstration of $\rm GVm^{-1}$ wakefields \cite{CPA-Laser-Wakefield-Expt,elec-Beam-Wakefield-Expt,AWAKE-nature}.

Tube-like but partially hollow gaseous plasma structures are considered desirable to counter beam ion interaction effects \cite{hollow-channel-beam} as well as to guide a focussed optical laser for laser wakefield electron acceleration \cite{laser-heated-capillary}. However, formation of structures in gaseous plasmas against its natural tendency has proven extremely difficult.

Nanofabrication offers the crucial advantage of atomic scale structural design. A nanofabricated near ideal hollow tube with tunable hundreds of nanometer core radius, $r_t$ and effective wall density, $n_t$ is introduced to sustain novel wakefields where a significant fraction of the oscillating tube wall electrons crunches into the core \cite{crunch-in-regime,crunch-in-regime-ipac,pre-signature-crunch}. These crunch-in nonlinear surface wave wakefields make possible the excitation of wall density wavebreaking fields ($E_{wb}[n_t]$). However, the solid-state crunch-in mode is yet to be modeled and understood.

Proof of principle of the crunch-in tube wakefield mechanism is established using 3D Particle-In-Cell (PIC) simulations. In Fig.\ref{fig:3D-wakefields-beam-tube}(a,b) the crunch-in tube surface mode driven as wakefield of an electron beam is evident from the 3D PIC electron density. The ionic lattice is stationary over tens of electron oscillations and the particle density is initialized to be zero within the tube core, $\lvert r \rvert <  r_t$. Longitudinal fields in excess of $10 \rm ~ TVm^{-1}$ ($10 \rm ~ TVm^{-1}$ wall focusing fields, see below) are evident in Fig.\ref{fig:3D-wakefields-beam-tube}(c) ($E_{wb}[n_t = 2\times\rm10^{22} cm^{-3}] = 13.6 ~ TVm^{-1}$).

The 3D simulations in Fig.\ref{fig:3D-wakefields-beam-tube} are carried out with \textsc{epoch} code \cite{epoch-pic} which incorporates quantum electrodynamics (QED) effects. A $\rm 3.6 \times 1.52 \times 1.52 ~ \mu m^3$ cartesian box with 2nm cubic cells is setup. The electrons in the tube of wall density, $n_t = \rm 2 \times 10^{22} cm^{-3}$ are modeled using 4 particles per cell with fixed ions. The tube has a radius, $r_t = \rm 100nm$ and wall thickness, $\Delta w \rm = 250nm$. An electron beam of peak density $n_{b0} = \rm 5\times 10^{21} cm^{-3}$; waist-size $\sigma_r = \rm 250 nm$ and bunch length $\sigma_z = \rm 400nm$ is initialized with 1 particle per cell. The box copropagates with this ultrarelativistic beam, $\gamma_b=10^4$. Absorbing boundary conditions are used for both fields and particles.

%%%%%%%%%%%%%%%%%%%%%%%%%%%%%%%%%%%%%%%%%
% FIGURE - Energy gain in Tube wakefield
%%%%%%%%%%%%%%%%%%%%%%%%%%%%%%%%%%%%%%%%%
\begin{figure}[!htb]
\centering
   \includegraphics[width=0.95\columnwidth]{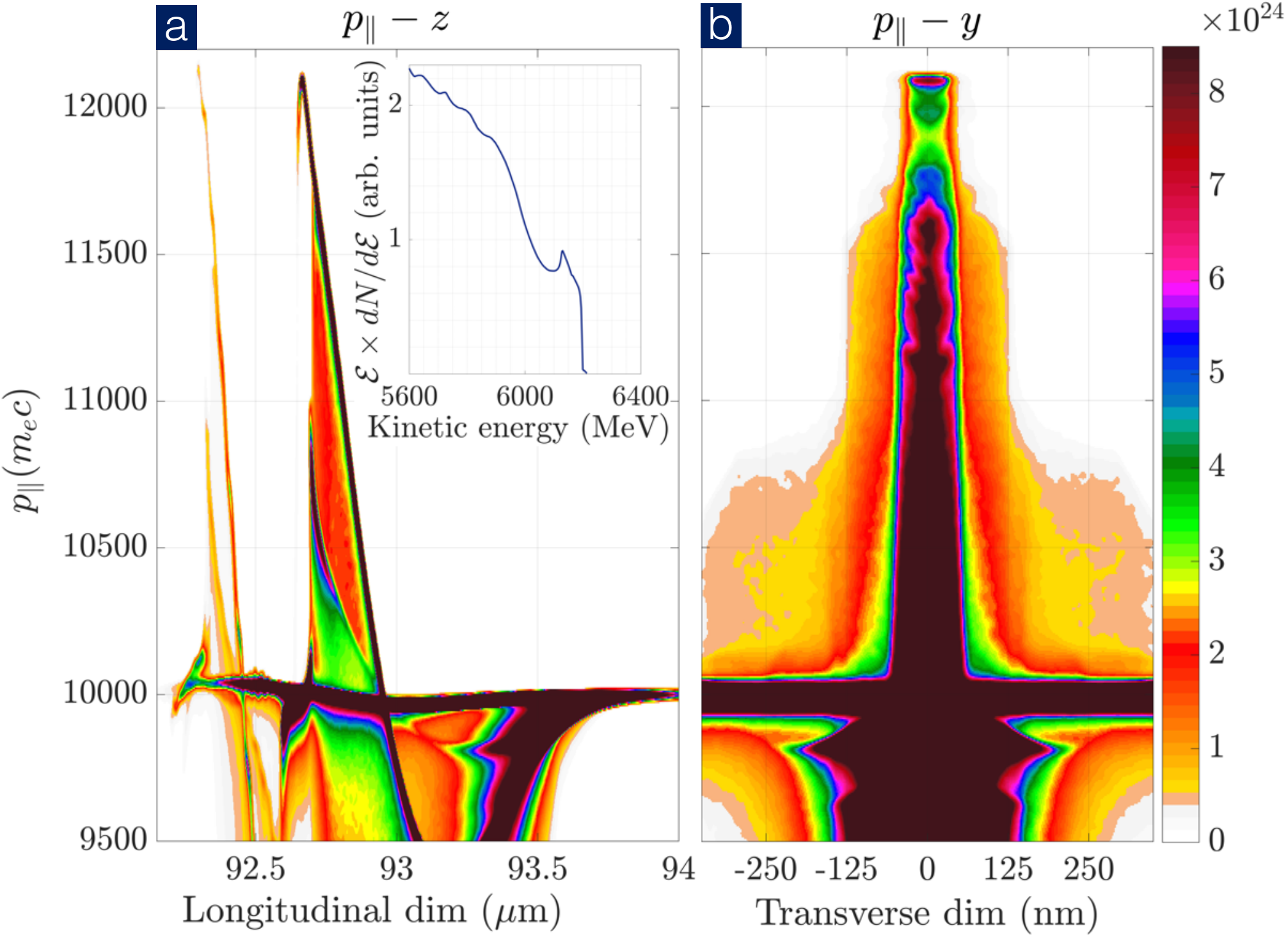}
   \caption{3D PIC simulation beam phase-spaces (a) $p_{\parallel} - z$ (energy spectrum inset) (b) $p_{\parallel} - y$ after $\sim\rm 93\mu m$ of interaction with the crunch-in tube wakefields of Fig.\ref{fig:3D-wakefields-beam-tube}.}
\label{fig:3D-phase-spaces-beam-tube}
\end{figure}

Acceleration of a bunch in the tail of the beam is demonstrated by these 3D simulations. Energy gain of 1.1GeV in $\rm 93\mu m$ long tube is inferred from the beam longitudinal momentum phase-spaces in Fig.\ref{fig:3D-phase-spaces-beam-tube} along the (a) longitudinal, (b) transverse dimensions. An average acceleration gradient of $\rm 11.6 ~ TVm^{-1}$ is obtained. The accelerated energy spectra inset in (a) is unoptimized because the realistic beam used in this proof of principle loads the entire range of acceleration phase.

The crunch-in tube wakefield mode in Fig.\ref{fig:3D-wakefields-beam-tube} is analytically modeled using collisionless kinetic theory. A charged particle beam propagates in the z-direction at $c\beta_b$ with a density profile, $n_b(r,z) = n_{b0} ~ \mathcal{F}(r,z)$ initially Gaussian shaped $\mathcal{F}(r,z) = {\rm exp}\left({-\frac{r^2}{2\sigma_r^2}}\right) ~ {\rm exp}\left({-\frac{(z-z_{\rm max})^2}{2\sigma_z^2}}\right)$, peak density $n_{b0} = \frac{N_b}{(2\pi)^{3/2} \sigma_r^2 \sigma_z}$ and $N_b =  \int_{-\infty}^{\infty} \int_0^{\infty} \int_0^{2\pi}  n_b(r,z) ~ d\theta ~ r dr ~ dz$ particles. The beam is sufficiently relativistic, $\gamma_b\gg 1$ such that its electric field is predominantly radial.

A necessary condition for the existence of tube wakefields is the mitigation of ``blowout''. Blowout drives a net momentum flux of all the tube electrons, $\Delta p(r)$ such that they altogether escape the restoring force of the tube ionic lattice. As an extreme case, all the tube electrons within an infinitesimal slice with net charge, $-e ~ n_t ~ \pi\left[ (r_t + \Delta w)^2 - r_t^2 \right]  ~ dz$ may bunch together and pile up into a compression layer just outside the outer tube wall, $r_t + \Delta w$. The net force on this layer is $(F_{\rm beam} + F_{\rm ion}) \Delta t = \Delta p$. If the outward force due to the beam, $F_{\rm beam}$ exceeds the restoring force of tube ions, $F_{\rm ion}$ then blowout occurs (see supplementary material). The tube and beam parameters thus have to satisfy the crunch-in condition, $\Delta p < 0$,
\begin{align}\label{eq:crunch-in-condition}
n_t\pi\left[ (r_t + \Delta w)^2 - r_t^2 \right] >  n_{b0} ~ \sigma_r^2
\end{align}

In the above 3D model the ratio of the left over right hand side of eq.\ref{eq:crunch-in-condition} is greater than 20. It may however be critical to optimize $\Delta w$ for considerations such as optimal wakefield spatial profile, vacuum etc.
 
The crunch-in kinetic model defines: $r_0$ as the equilibrium position of a tube electron with, $r_t < r_0 < r_t+\Delta w$; $r(z,t)$ as the instantaneous radial position of an oscillating tube electron; $r_{\rm max}$ as the maximum radius where the driven tube electrons form a compression layer; $\mathcal{H}$ as the step function with $\mathcal H(0^+) = 1$ and $\mathcal H(0^-) = 0$ to model the effect of step transition in tube wall density.

The tube electrons which are located at an equilibrium radius less than the electron under consideration at $r_0$ (between $r_0$ and $r_t$) also collectively move with it and compress at the radial extrema. When all the tube electrons with an equilibrium radii between $r_t$ and $r_0$ move together to form a compression layer at a new radial location $r$, the ionic force on the electron under consideration is $-e ~ E_{\rm ion}(r > r_t) = - 4\pi e^2 n_t ~ \frac{(r^2 - r_t^2)}{2r}$.

In addition to the ionic force, all the electrons with an equilibrium radii smaller than $r_0$ which collectively move with the electron under consideration result in a collective field opposite to the ionic field. The collective oscillation condition requires that the electrons that originate at an equilibrium position less than $r_0$ collective move to a position just behind $r$. The force due to collectively moving tube electrons located between $r_0$ and $r_t$ is thus $-e ~ E_{\rm e}(r) = 4\pi e^2 n_t ~ \frac{(r_0^2 - r_t^2)}{2r}$.
 
The dynamics is different during the radially inward moving phase of the oscillation. Due to the zero ion density in the core region of the tube, the collectively moving electrons do not experience any ionic force. However, collectively moving tube electrons that originate between $r_0$ and $r_t$ crunch into the tube core. Inside the core an increase in mutual electrostatic field of the compressing electrons forces them back towards equilibrium. Thus, the net force acting on the electrons is, $F_{\rm collective} = - m_e ~ \frac{\omega_{pe}^2(n_t)}{2} ~ \frac{1}{r} ~ \left[ (r^2 - r_t^2) ~ \mathcal{H}(r-r_t) - (r_0^2 - r_t^2)  \right]$. 

When the tube electrons are driven by an electron (positron) beam with ${\rm sgn}[Q_b] = -1$ ($+1$), they are initially pushed radially outwards (inwards). The acceleration of an oscillating electron as it traverses across the the density discontinuity at the inner surface $r = r_t$ is thus considered, $\frac{d^2 r}{dt^2} |_{r=r_t} ~ \mathcal{H}({\rm sgn}[Q_b](r-r_t ))$.

The equation of relativistic ($\gamma_e$) collective surface electron oscillation is $ \gamma_e m_e ~ \frac{\partial^2 r(\xi,z,t)}{\partial t^2} = F_{\rm collective}$. Equation of crunch-in surface wave is obtained by transforming to a frame $\xi = c\beta_b t - z$, co-moving with the driver ($\beta_b$ for an x-ray laser is its group velocity) and using $\partial \xi= ({c\beta_b})^{-1} \partial t$. By including the force of the ultrashort drive bunch the driven crunch-in surface wave equation is,
\begin{align}\label{eq:surface-wave-equation}
%\begin{split}
\nonumber \frac{\partial^2 r}{\partial \xi^2} & + \frac{1}{2}\frac{\omega_{pe}^2(n_t)}{\gamma_e c^2\beta_b^2} ~ \frac{1}{r} ~ \left[ (r^2 - r_t^2) ~ \mathcal{H}(r-r_t) - (r_0^2 - r_t^2)  \right] \\
\nonumber & + \frac{\partial^2 r}{\partial \xi^2} \biggm\lvert_{r=r_t} \mathcal{H}({\rm sgn}[Q_b](r-r_t )) \\ 
& = - {\rm sgn}[Q_b] ~ \frac{\omega_{pe}^2(n_t)}{\gamma_e c^2\beta_b^2} ~ \frac{n_{b0}(\xi)}{n_t} ~ \int_{0}^{r} dr ~ \mathcal{F}(r,z,\xi)
%~ \frac{\sigma_r^2}{2\pi} ~ \frac{\left[ 1 - {\rm exp}\left( -\frac{r^2}{2\sigma_r^2} \right) \right] }{r} ~ {\rm exp}\left({-\frac{(z-z_{\rm max})^2}{2\sigma_z^2}}\right)
%\end{split}
\end{align}
%%%%%%%%%%%%%%%%%%%%%%%%%%%%%%%%%%%%%%%%%%%%%%%%%%%%%%%%%%%%%%%%%%%%%
% SECTION - Discussion and Future Work
%%%%%%%%%%%%%%%%%%%%%%%%%%%%%%%%%%%%%%%%%%%%%%%%%%%%%%%%%%%%%%%%%%nd%
%\section{Discussion and Future Work}
%%%%%%%%%%%%%%%%%%%%%%%%%%%%%%%%%%%%%%%%%%%%%%%%%%%%%%%%%%%%%%%%%%%%%
% SECTION - 3D XLAS driven mode
%%%%%%%%%%%%%%%%%%%%%%%%%%%%%%%%%%%%%%%%%%%%%%%%%%%%%%%%%%%%%%%%%%nd%
%\section{3D XLAS driven mode}
With a Gaussian beam envelope, $ \int_{0}^{r} dr ~ \mathcal{F}(r,z,\xi) = \frac{\sigma_r^2}{2\pi} ~ \left[ 1 - {\rm exp}\left( -\frac{r^2}{2\sigma_r^2} \right) \right] ~ r^{-1} ~ {\rm exp}\left({-\frac{(z-z_{\rm max})^2}{2\sigma_z^2}}\right)$ and under flat-top condition, $\sigma_r\gg r_t$ maxima of the radial trajectory, $r=r_m$ is obtained from eq.\ref{eq:surface-wave-equation}. At this maxima, force of the drive beam equals that of the collective charge separation field. The electrons located between $r_m$ and $r_t$ collectively move and bunch into a compression layer located at $r_m = r_t \quad \left( 1 - \frac{n_{b0}}{n_t} ~ \frac{2\pi\sigma_r^2}{\pi(r_t+\Delta w)^2}  \right)^{-1/2}$.

The net charge of the electron compression layer that collectively crunches from the tube wall into its core can be estimated. During this crunch-in phase the displaced electrons fall into the core region up to a minimum radius, $r_{\rm min}$. The net charge that falls into the core region is $\delta Q_{\rm max}(r_{\rm min}) = - e ~ n_t ~ \pi r_t^2  ~ \left( \frac{n_{b0} ~ 2\pi\sigma_r^2}{n_t \pi (r_t + \Delta w)^2 - n_{b0} ~ 2\pi\sigma_r^2 }  \right) ~ dz$.

The radial electric field is thus obtainable applying the Gauss's law on $\delta Q_{\rm max}$. Although an analytical expression for $r_{\rm min}$ can be obtained, we consider $r_{\rm min} = r_t / \alpha$. The radial electric field is  $E_{t-r} = - \alpha ~ n_t ~ 2\pi r_t  ~ \left( \frac{ e n_{b0} ~ 2\pi\sigma_r^2}{n_t \pi (r_t + \Delta w)^2 - n_{b0} ~ 2\pi\sigma_r^2 }  \right)$ which simplifies to,
\begin{align}\label{eq:crunch-in-charge-radial-field-SI}
\begin{split}
E_{t-r} = & - \alpha ~ \sqrt{\frac{2}{\pi}} ~ \frac{Q_b[{\rm pC}]}{\sigma_z{\rm [100nm]}} ~  \frac{1}{r_t[100{\rm nm}]}  \\ & \times ~ \left(  \left( \frac{r_t + \Delta w}{r_t} \right)^2  - 2 \frac{n_{\rm b0}}{n_t}\frac{\sigma_r^2}{r_t^2} \right)^{-1} ~ {\rm \frac{TV}{m}}
\end{split}
\end{align}

%%%%%%%%%%%%%%%%%%%%%%%%%%%%%%%%%%%%%%%%%
% FIGURE - Focusing field and beam envelope modulation
%%%%%%%%%%%%%%%%%%%%%%%%%%%%%%%%%%%%%%%%%
\begin{figure}[!htb]
\centering
   \includegraphics[width=0.75\columnwidth]{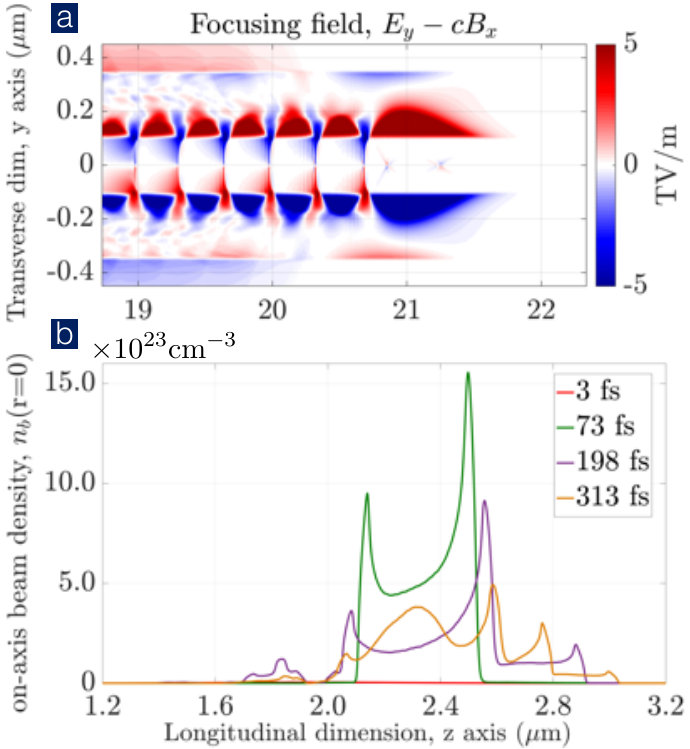}
   \caption{3D PIC (a) tube focusing wakefield, (b) on-axis beam density which demonstrates the nanomodulation effect.}
\label{fig:3D-focusing-beam-nanomod}
\end{figure}

The peak longitudinal electric field is derived using the Panofsky-Wenzel theorem, $E_{t-r} ~ \Delta r = E_{t-z} ~ \Delta \xi$. The value of $E_{t-z}$ varies over $\kappa ~ \sqrt{\gamma_e} ~ 2\pi c/\omega_{pe}(n_t)$ where $\kappa$ is the shortened phase of the nonlinearly steepened surface wave, where the tube electrons crunch into its core. The relativistic factor, $\gamma_e$ ($\simeq (1 + (p_r^2/(m_ec)^2)^{1/2}$) reduces the oscillation frequency as $\omega_{pe}(n_t)/\sqrt{\gamma_e}$. Using $p_r = F_{\rm beam} ~ \sigma_z/c  = 4\pi e^2 n_tc^{-1} ~ n_{b0}n_t^{-1} ~ r_t \sigma_z(4\pi)^{-1}$, the peak longitudinal field is thus, $E_{t-z} = E_{t-r} ~ r_{\rm min}(\kappa ~ 2\pi c)^{-1} ~ \omega_{pe}(n_t)\gamma_e^{-1/2}$ or,
\begin{align}\label{eq:crunch-in-charge-acc-field-SI}
\begin{split}
E_{t-z} = & - \frac{2}{\kappa} ~ \frac{ \sqrt{n_t[10^{22}{\rm cm^{-3}}] } }{{\sqrt{ r_t \rm [100nm]} }} ~ \sqrt{Q_b[{\rm pC}]} ~ \frac{\sigma_r}{\sigma_z} \\
& \times \left(  \left( \frac{r_t + \Delta w}{r_t} \right)^2  - 2 \frac{n_{\rm b0}}{n_t}\frac{\sigma_r^2}{r_t^2} \right)^{-1} ~ {\rm \frac{TV}{m}} 
\end{split}
\end{align}

The expressions of tube wakefield in eq.\ref{eq:crunch-in-charge-radial-field-SI} and eq.\ref{eq:crunch-in-charge-acc-field-SI} are applicable only if the crunch-in condition in eq.\ref{eq:crunch-in-condition} is strictly satisfied. Moreover, closer to the critical point in eq.\ref{eq:crunch-in-condition} the wakefield amplitudes depend on the ratio.

For $n_t = 2\times 10^{22} {\rm cm^{-3}}$, $r_t = 100 {\rm nm}$ and $Q_b = 315 ~ {\rm pC}$, $\sigma_z=250{\rm nm}$, $\sigma_z=400{\rm nm}$; $r_m=74.5\rm nm$, $E_{t-r} = \alpha ~ 8.96 ~ {\rm TV/m}$ (eq.\ref{eq:crunch-in-charge-radial-field-SI}) and $E_{t-z} = \kappa^{-1} ~ 3.5 ~ {\rm TV/m}$ (eq.\ref{eq:crunch-in-charge-acc-field-SI}) in good agreement with the above 3D simulation.

The beam envelope in eq.\ref{eq:surface-wave-equation} is considered to be quasi-stationary over several surface oscillations. Transverse  envelope oscillations however result in the variation of beam spatial profile, $\mathcal{F}(r,z,\xi)$ and peak density, $n_{b0}(\xi)$. 

%A kinetic equation of the radial dynamics of a beam particle at $r_b(\xi,t)$ under the two forces from beam self-fields and the tube wall focusing fields (in the region of the bared tube wall ions) is thus obtained in the beam frame.
%\begin{align}\label{eq:beam-envelope-equation}
%\nonumber & \frac{\partial^2 r_b}{\partial \xi^2} + \frac{1}{2}\frac{\omega_{pe}^2(n_t)}{\gamma_e c^2\beta_b^2} ~ \frac{(r_b^2 - r_t^2)}{r_b} ~ \mathcal{H}(r_b - r_t) \mathcal{H}(r_m - r_b) \\
%& \qquad = - {\rm sgn}[Q_b] ~ \frac{\omega_{pe}^2(n_t)}{\gamma_e c^2\beta_b^2} ~ \frac{n_{b0}(\xi)}{n_t} ~ \int_{0}^{r_b} dr ~ \mathcal{F}(r,z,\xi)
%\end{align}

The tube focusing fields and nanometric transverse beam oscillations from the above 3D simulation are elucidated in Fig.\ref{fig:3D-focusing-beam-nanomod}. The beam particles within $r_m > r_b > r_t$ experience transverse focusing and the beam electrons are forced into the core which results in the folding in of the ``wings'' of the beam. The beam develops significant nano modulation with spatial frequencies corresponding to $\lambda_{\rm osc}\sim\rm \mathcal{O}(100nm)$ as shown in Fig.\ref{fig:3D-focusing-beam-nanomod}(b). Moreover, the peak on-axis beam density $n_{b0}(\xi)$ rises to many ten times its initial density, $\rm n_{b0}(\xi=0)=5\times 10^{21}cm^{-3}$. With this rapid rise in the beam density the tube wakefield amplitude approaches the wavebreaking limit.

The $\rm \mathcal{O}(100MeV)$ high-energy radiation ($\textsc{e}_{\rm ph}=hc ~ 2\gamma_b^2/\lambda_{\rm osc}$) produced by nanometric oscillations of ultrarelativistic particles of the beam in the $10 \rm ~ TVm^{-1}$ wall focusing fields from a nanometric source size ($\sim  r_t$) offers a nano-wiggler photon source. Furthermore, variation of tube wall density (nanolattice) or inner radius (corrugated nanostructure) can be used to further enhance the beam oscillations and thus the radiation characteristics. Our future work will investigate schemes to accentuate the collective nature of beam nanomodulation.

%%%%%%%%%%%%%%%%%%%%%%%%%%%%%%%%%%%%%%%%%
% FIGURE - x-ray laser driven Tube wakefield
%%%%%%%%%%%%%%%%%%%%%%%%%%%%%%%%%%%%%%%%%
\begin{figure}[!htb]
\centering
   \includegraphics[width=\columnwidth]{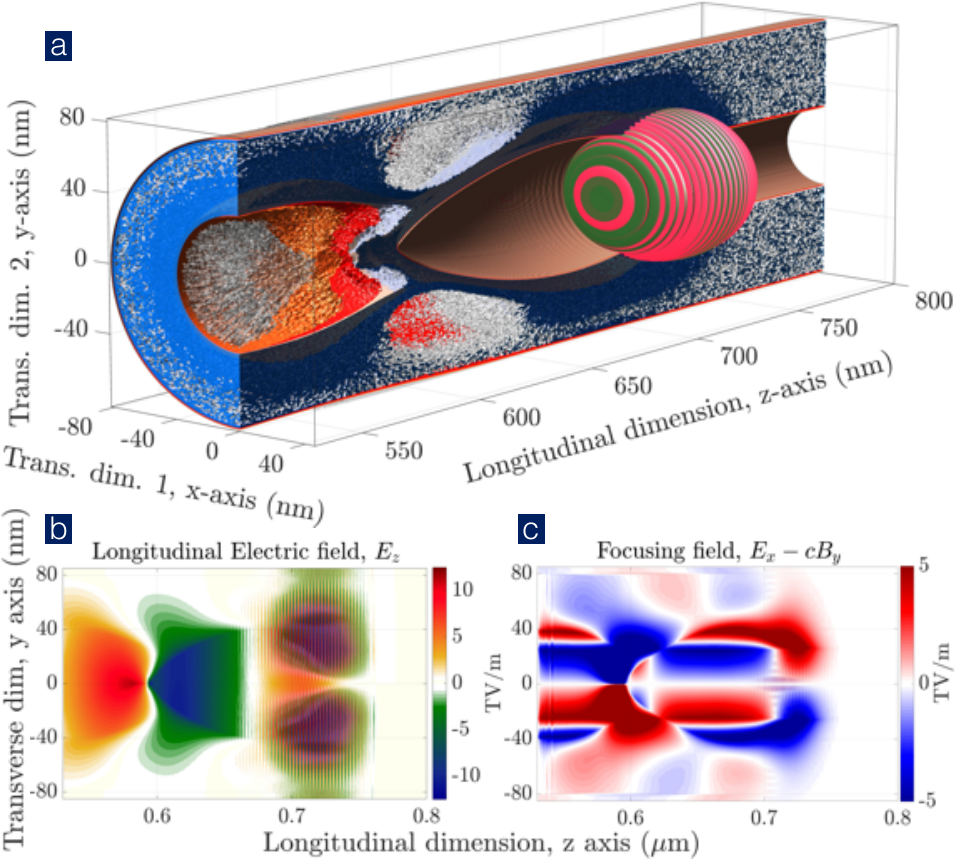}
   \caption{3D PIC simulations of nano x-ray laser driven tube wakefield (a) electron density, (b) acceleration and (c) focusing field profile at around $\rm 1\mu m$ (3fs) of interaction.}
\label{fig:3D-wakefields-xlas-tube}
\end{figure}

Although a nano x-ray laser is yet to be prototyped, 3D simulation in Fig.\ref{fig:3D-wakefields-xlas-tube} demonstrates that x-ray pulse driven crunch-in mode is almost identical to above. In Fig.\ref{fig:3D-wakefields-xlas-tube} a 2.5mJ x-ray pulse is initialized with $a_0=2.7$ in line with the conceptual design in \cite{xray-Mourou} with photon energy 500\rm eV ($\lambda_0\rm=2.5nm$), pulse length $\tau_{\rm FWHM} = 150 ~ \text{attosec}$ and waist-size $w_{\rm FWHM} = \rm 25nm$. The interaction is modeled in a $\rm 270 \times 170 \times 170 ~ nm^3$ cartesian box with $\rm 1.25 \times 1.67 \times 1.67 ~ \AA^3$ cells. The tube parameters are $n_t=\rm 7.0\times10^{22}cm^{-3}$, $r_t\rm =25nm$ and $\Delta w \rm = 55nm$. The longitudinal wakefield in Fig.\ref{fig:3D-wakefields-xlas-tube}(b) is $E_{t-z} \simeq \rm 12.5 ~ TVm^{-1}$ ($E_{wb} = \rm 25.4 ~ TVm^{-1}$). The focusing field in Fig.\ref{fig:3D-wakefields-xlas-tube}(c) however differs from Fig.\ref{fig:3D-focusing-beam-nanomod}(a).

%In conclusion, the crunch-in regime of nonlinear surface wave driven as wakefield in nanostructured tubes makes a nano wakefield accelerator realizable. Moreover, the $\rm TVm^{-1}$ wall focusing fields can be used to drive nanometric oscillation of an ultrarelativistic beam for controlled photon production. Therefore, the atomic scale $\rm TVm^{-1}$ wakefields modeled here open up unprecedented pathways for acceleration and radiation generation.

In conclusion, our 3D PIC and analytical modeling demonstrates that near solid submicron multi-GeV electron bunch can effectively excite $\rm\mathcal{O}(TVm^{-1})$ longitudinal crunch-in wakefields in nanostructures, such as a tube of 200nm core diameter. We discover that nonlinear surface crunch-in waves also sustain many $\rm TVm^{-1}$ focusing wakefields in the tube walls which result in more than an order of magnitude increase in the peak beam density and hundred nm electron beam density modulation. Even with currently available electron bunches and nanofabrication technologies, the resulting accelerating fields can reach unprecedentedly high levels to allow the demonstration of $\rm\mathcal{O}(GeV)$ energy gains in mm long tubes. This may not be matched by any other advanced acceleration concept. Besides the unmatched rapid particle acceleration, induced nanomodulation also opens up controlled $\rm\mathcal{O}(100MeV)$ radiation production using the nanometric transverse oscillations of the beam particles. A nano wakefield accelerator may thus help open new frontiers in a wide-range of scientific areas.

%%%%%%%%%%%%%%%%%%%%%%%%%%%%%%%%%%%%%%%%%%%%%%%%%%%%%%%%%%%%%%%%%%%%%
% SECTION - Acknowledgement
%%%%%%%%%%%%%%%%%%%%%%%%%%%%%%%%%%%%%%%%%%%%%%%%%%%%%%%%%%%%%%%%%%%%%
\begin{acknowledgments}
A. A. S. was supported by the College of Engineering, University of Colorado Denver.
T. T. was partially supported by the Rostoker fund at the University of California Irvine.
V. D. S. was supported by Fermilab operated by the Fermi Research Alliance, LLC under Contract No. DE-AC02-07CH11359 with the US DoE.
We acknowledge discussions \cite{crystal-workshops} at the ACN 2020 workshop organized by CERN-ARIES (Mar 2020), ATS seminar at CERN (Feb 2020), FACET-II Science Review meeting (Oct 2019) and XTALS 2019 workshop at Fermilab (Jun 2019). 
We also acknowledge inputs from Prof. P. Taborek on nanofabrication and nanomaterial characterization and Prof. G. Mourou on x-ray lasers. 
The NSF XSEDE RMACC Summit supercomputer utilized here was supported by the NSF awards ACI-1548562, ACI-1532235 and ACI-1532236, the Univ. of Colorado Boulder, and Colorado State Univ. \cite{xsede-rmacc-citation}.
\end{acknowledgments}

%%%%%%%%%%%%%%%%%%%%%%%%%%%%%%%%%%
% REFERENCES
%%%%%%%%%%%%%%%%%%%%%%%%%%%%%%%%%%

\end{document}